\documentclass{ws-ijmpd}
\usepackage[super,compress]{cite}
\begin{document}

\markboth{Ksh. Newton Singh {\it et al.}}
{A new solution of  embedding class I representing anisotripic fluid sphere in general relativity}

\catchline{}{}{}{}{}
%
%%%%%%%%%%%%%%%%%%%%%%%%%%%%%%%%%%%%%%%%%%%%%%%%%%%%%%%%%%%%%%%%%%%%

\title{A new solution of  embedding class I representing anisotripic fluid sphere in general relativity}

\author{Ksh. Newton Singh}

\address{Department of Physics, National Defence Academy\\
 Khadakwasla, Pune, Maharashtra-411023, India\\
ntnphy@gmail.com}

\author{Piyali Bhar}

\address{Department of Mathematics, Government General Degree College, Singur,\\
Hooghly, West Bengal-712409, India\\
piyalibhar90@gmail.com}

\author{Neeraj Pant}

\address{Department of Mathematics, National Defence Academy\\
 Khadakwasla, Pune, Maharashtra-411023, India\\
neeraj.pant@yahoo.com}

\maketitle

\begin{history}
\received{Day Month Year}
\revised{Day Month Year}
\end{history}

\begin{abstract}
In the present paper we are willing to model anisotropic star by choosing a new $g_{rr}$ metric potential. All the physical parameters like the matter density, radial and transverse pressure and are regular inside the anisotropic star, with the speed of sound less than the speed of light. So the new solution obtained by us gives satisfactory description of realistic astrophysical compact stars. The model of the present paper is compatible with observational data of compact objects like RX J1856-37, Her X-1, Vela X-12 and Cen X-3. A particular model of Her X-1 (Mass 0.98 $M_{\odot}$ and radius=6.7 km.) is studied in detail and found that it satisfies all the condition needed for physically acceptable model. Our model is described analytically as well as with the help of graphical representation.
\end{abstract}


\ccode{PACS numbers: 04.20.-q; 04.40.Nr; 04.40.Dg}

\section{Introduction}
To find the exact solution of Einstein field equations is always an interesting topic to the researchers. In 1916 the first exact solution of Einstein field equations for the interior of a compact object was obtained by Schwarzschild and later several relativists obtained new exact solutions. There are a large number of works in literature based on the topic of the exact solutions but very few of them satisfy the required general physical conditions inside the stellar interior. From the analysis of Delgaty \& Lake \cite{lake88} it is well known that out of 127 published solutions only 16 solutions satisfy all the physical conditions. Newton Singh \& Pant \cite{ntn1,ntn2,ntn3} have given a class of exact solutions of Einstein's gravitational field equations describing spherically symmetric and static anisotropic stellar type configurations. The solutions were obtained by assuming a particular form of the anisotropy factor. Fodor \cite{foder00} has proposed an algorithm to generate any number of physically realistic pressure and density profiles for spherical perfect fluid isotropic distributions without evaluating integrals. Mak, Dobson \& Harko\cite{mak00} have derived the upper limits for the mass-radius ratio for compact general relativistic objects in presence of cosmological constant as well as in presence of a charged distribution\cite{mak01} . By assuming a particular mass function Sharma \& Maharaj \cite{sharma07} found new exact solutions to the Einstein field equations with an anisotropic matter distribution. A distinguishing feature of this class of solutions is that they admit a linear equation of state which can be applied to strange stars with quark matter. Tolman IV solution in the Randall-Sundrum Braneworld was obtained by Ovalle \& Linares \cite{Ovalle13} . In the context of the Randall-Sundrum braneworld, the minimal geometric deformation approach (MGD) is used to generate an exact analytic interior solution to four-dimensional effective Einstein's field equations for a spherically symmetric compact distribution. By using this analytic solution, the authors have developed an exhaustive analysis of the braneworld effects on realistic stellar interiors, finding strong evidences in favor of the hypothesis they have also shown that compactness is reduced due to bulk effects on stellar configurations.\\

The study of spherically symmetric solution to Einstein equations allow for finding the anisotropy pressures (the radial component of the pressure differs from the angular component)in relativistic astrophysics, which got much attention after the extensive investigations by Bowers \& Liang\cite{Bowers} , and the theoretical investigations by Ruderman\cite{Ruderman} . Anisotropy of principal pressures can be found when the source is derived from field theories, i.e., when scalar fields are considered as in boson stars \cite{Schunck} , and the role of strange matter with densities higher than neutron stars ($\sim 10^{15}$ gm/cc).\\

Local anisotropy in self-gravitating systems were studied by Herrera and Santos \cite{herrera97} and see the the references there in for a review of anisotropic fluid sphere. Anisotropy may occurs in various reasons e.g, the existence of solid core, in presence of type P superfluid, phase transition, rotation, magnetic field, mixture of two fluid, existence of external field etc. To create pressure anisotropy different mechanisms in stellar models have been identified by Ivanov \cite{iva} . Durgapal and Bannerji\cite{dur} proposed a model of neutron star satisfying all the physical requirements. A good collections of exact solution of Einstein's field equation can be found in \cite{step,chaisi,dev1,dev2,herrera1,herrera2,ivanov,mak1}. Li {\em et al.}\cite{li} investigated the structure and stability properties of compact astrophysical objects that may be formed from the Bose-Einstein condensation of dark matter. They also studied numerically the structure equations of the condensate dark matter stars.\\

B\"{o}hmer and Harko \cite{harko1} derived upper and lower limits for the basic physical parameters $viz.$ mass-radius ratio, anisotropy, redshift and total energy for arbitrary anisotropic general relativistic matter distributions in the presence of cosmological constant. They have shown that anisotropic compact stellar type objects can be much more compact than the isotropic ones, and their radii may be close to their corresponding Schwarzschild radii. A new model of anisotropic compact star by assuming Tolman VII gravitational potential for $g_{rr}$ metric has been obtained by Bhar {\em et al.}\cite{murad15} . They obtained an anisotropic compact star of mass $0.41M_{\odot}$, radius 3.8 km. and central density $3.98\times 10^{15}$ gm/cc. In a recent paper Bhar \cite{bhar15a} obtained a model of compact star admitting chaplygin equation of state in Finch-Skea spacetime of radius 9.69 km and the mass to be $2.04M_{\odot}$, which is very close to the observational data of the strange star PSR J1614-2230 reported by Gangopadhyay {\em et al.} \cite{gan} . Many articles have also discussed on embedding class I solutions that can represent compact stars \cite{mau,gup1,gup2} .

% organization of the paper
The aim of this paper is to generate a new model of anisotropic relativistic anisotropic star satisfying the Karmakar's\cite{kar} condition. In Section 2, we describe the basic field equations in the form of differential equations governing by the gravitational field and the condition which should satisfy to represent a model of anisotropic compact star of embedding class I. A new model of anisotropic compact star has been obtained in Section 3 by assuming a new form for $g_{rr}$. The values of the integration constants $A$ and $B$ are obtained from the matching condition which is given in Section 4. In next section we discuss about the physical analysis of our present model. In the final section we summarize all the obtained results and have shown that the model presented here may relate to various compact stars presented in Table 1 and 2.

\section{Interior spacetime and the Einstein Field Equations}

The interior of the super-dense star is assumed to be described by the line element
\begin{equation}
ds^2 = e^{\nu(r)}dt^2 - e^{\lambda(r)}dr^2 - r^2(d\theta^2 + \sin^2{\theta}d\phi^2) \label{metric}
\end{equation}
Where $\nu$ and $\lambda$ are functions of the radial coordinate `$r$' only.

Let us assume that the matter within the star is anisotropic in nature and correspondingly the energy-momentum tensor is described by,
\begin{equation}
T_{\nu}^{\mu}=(\rho+p_r)u^{\mu}u_{\nu}-p_t g_{\nu}^{\mu}+(p_r-p_t)\eta^{\mu}\eta_{\nu} \label{ten}
\end{equation}
with $ u^{i}u_{j} =-\eta^{i}\eta_j = 1 $ and $u^{i}\eta_j= 0$, the vector $u_i$ being the fluid 4-velocity and $\eta^{i}$ is the spacelike vector which is orthogonal to $ u^{i}$. Here $\rho$ is the matter density, $p_r$ is the the radial and $p_t$ is transverse pressure of the fluid in the orthogonal direction to $p_r$.\\

Now for the line element (\ref{metric}) and the matter distribution (\ref{ten}) Einstein Field equations (assuming $G=c=1$) is given by,
\begin{eqnarray}\label{g3}
8\pi \rho &=& \frac{\left(1 - e^{-\lambda}\right)}{r^2} + \frac{\lambda'e^{-\lambda}}{r} \label{g3a} \\ \nonumber \\
8\pi p_r &=&  \frac{\nu' e^{-\lambda}}{r} - \frac{\left(1 - e^{-\lambda}\right)}{r^2} \label{g3b}\\ 
8\pi p_t &=& \frac{e^{-\lambda}}{4}\left(2\nu'' + {\nu'}^2  - \nu'\lambda' + \frac{2\nu'}{r}-\frac{2\lambda'}{r}\right) \label{g3c}
\end{eqnarray}
where primes represent differentiation with respect to the radial coordinate $r$. Using Eqs. (\ref{g3b}) and (\ref{g3c}) we get
\begin{eqnarray}
8\pi\Delta &=& 8\pi (p_t-p_r) = e^{-\lambda}\left[{\nu'' \over 2}-{\lambda' \nu' \over 4}+{\nu'^2 \over 4}-{\nu'+\lambda' \over 2r}+{e^\lambda-1 \over r^2}\right]+{1\over r^2}\label{del}
\end{eqnarray}

\begin{figure}[!htb]\centering
   \begin{minipage}{0.5\textwidth}
    \includegraphics[width=\linewidth]{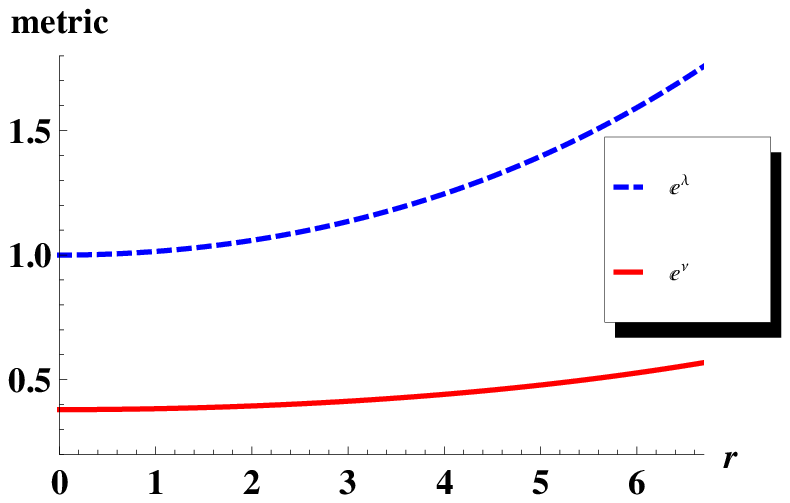}
     \end{minipage}
   \begin {minipage}{0.45\textwidth}
    \includegraphics[width=\linewidth]{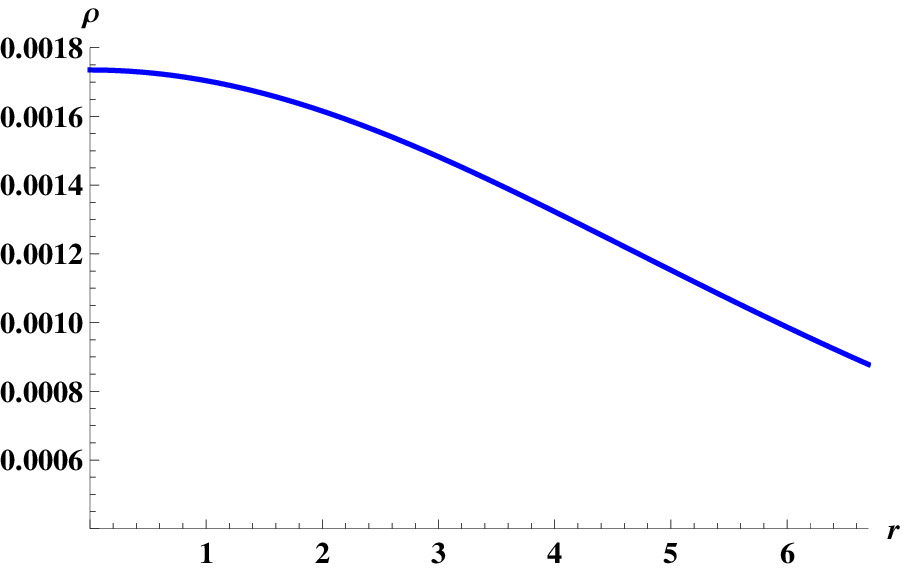}
     \end{minipage}
    \caption{(Left) The metric potentials are plotted against $r$ for the compact star Her X-1 by taking $b=007268209067$, $A=0.07395029811$ and $B=0.04902096388$. $e^{\lambda}$ is shown in dashed line (Blue color) and $e^{\nu}$ is shown in solid line (red color). (Right) Variation of matter density is plotted against $r$ for the compact star Her X-1. }\label{f1}
\end{figure}

If the metric given in (\ref{metric}) satisfies the Karmakar's \cite{kar} condition , it will represent an embedding class I spacetime i.e.
\begin{equation}
R_{1414}={R_{1212}R_{3434}+R_{1224}R_{1334} \over R_{2323}}\label{con}
\end{equation}
with $R_{2323}\neq 0$, \cite{pandey} . This condition leads to a differential equation given by
\begin{equation}
{2\nu'' \over \nu'}+\nu'={\lambda' e^\lambda \over e^\lambda-1}\label{dif1}
\end{equation}

\begin{figure}[!htb]\centering
   \begin{minipage}{0.5\textwidth}
     \includegraphics[width=\linewidth]{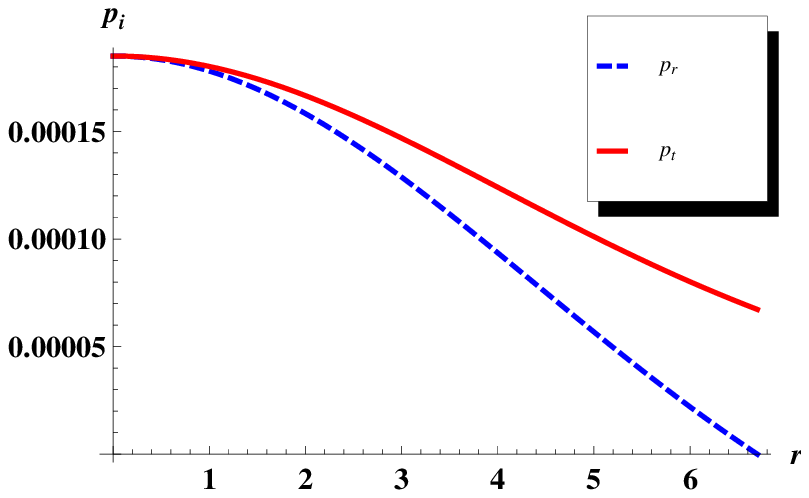}
        \end{minipage}
   \begin {minipage}{0.45\textwidth}
     \includegraphics[width=\linewidth]{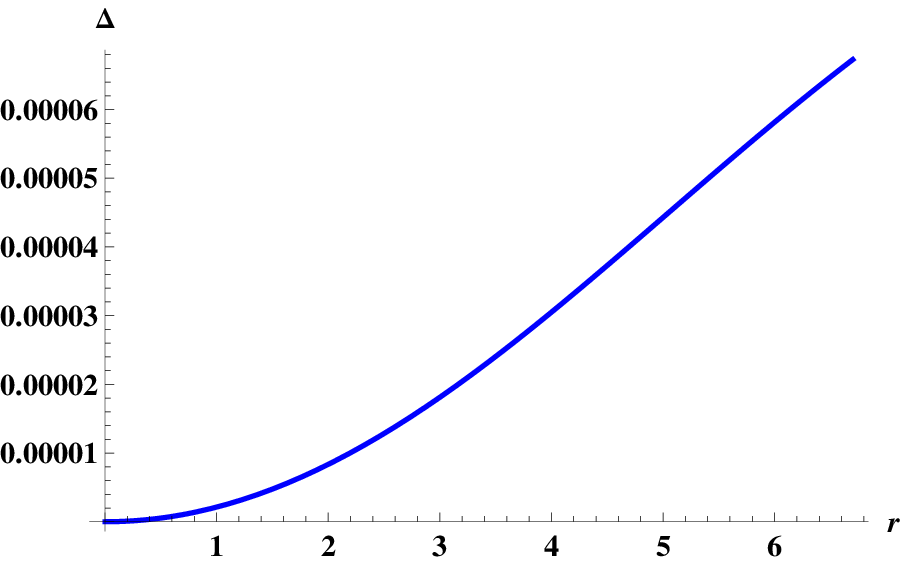}
        \end{minipage}
   \caption{(Left) The radial(dashed line in blue color) and transverse pressure (solid line in red color) are plotted against $r$ for the compact star Her X-1 by taking the same values of the constant mentioned in Fig. 1. (Right) Variation anisotropic factor is plotted against $r$ for the compact star Her X-1 by taking the same values of the constant mentioned in Fig. 1}\label{f2}
\end{figure}

On integration we get the relationship between $\nu$ and $\lambda$ as
\begin{equation}
e^{\nu}=\left(A+B\int \sqrt{e^{\lambda}-1}~dr\right)^2\label{nu1}
\end{equation}
where $A$ and $B$ are constants of integration.

By using (\ref{nu1}) we can rewrite (\ref{del}) as
\begin{eqnarray}
8\pi\Delta = {\nu' \over 4e^\lambda}\left[{2\over r}-{\lambda' \over e^\lambda-1}\right]~\left[{\nu' e^\nu \over 2rB^2}-1\right] \label{del1}
\end{eqnarray}
Here $\Delta=(p_t-p_r)$ is the measure of anisotropy, which will be attractive in nature if $p_t>p_r$ and repulsive if $p_t<p_r.$

\section{A new class of well-behaved embedding class-I solution}

To solve the above equation (\ref{nu1}), let us assume the metric potential $g_{rr}$ as follows:
\begin{equation}
e^\lambda =(1+b r^2)^2 \label{elam}
\end{equation}
where  $b$ is constant having a dimension of length$^{-2}$ and it will be obtained later from the matching conditions.

Using the metric potential (\ref{elam}) in (\ref{nu1}), we obtained the expression for the metric potential $e^{\nu}$ as,
\begin{eqnarray}
e^\nu=\left(A+\frac{B(b r^2+2)^{3/2}}{3 \sqrt{b}}\right)^2 \label{enu}
\end{eqnarray}

Using (\ref{elam}) and (\ref{enu}), we can rewrite the expression for matter density $\rho$, radial pressure $p_r$, anisotropic factor $\Delta$ and transverse pressure $p_t$ as
\begin{eqnarray}
8\pi \rho & = & \frac{b (b^2 r^4+3 b r^2+6)}{(b r^2+1)^3}\\
8\pi p_r & = & \frac{b}{(b r^2+1)^2}\left[\frac{6 B \sqrt{b r^2+2}}{3 A \sqrt{b}+B \left(b r^2+2\right)^{3/2}}-b r^2-2\right]\\
8\pi\Delta & = & \frac{b^2 r^2 \left(b r^2+3\right) \left(3 A \sqrt{b} \sqrt{b r^2+2}+b^2 B r^4+4 b B r^2+B\right)}{\left(b r^2+1\right)^3 \sqrt{b r^2+2} \left(3 A \sqrt{b}+B \left(b r^2+2\right)^{3/2}\right)}\\
p_t & = & p_r+\Delta
\end{eqnarray}

\begin{figure}[!htb]\centering
   \begin{minipage}{0.49\textwidth}
     \includegraphics[width=\linewidth]{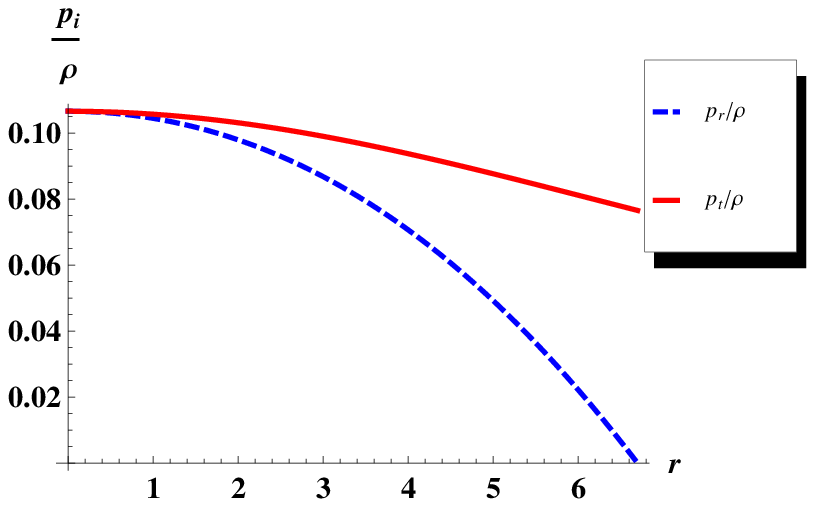}
        \end{minipage}
   \begin {minipage}{0.49\textwidth}
     \includegraphics[width=\linewidth]{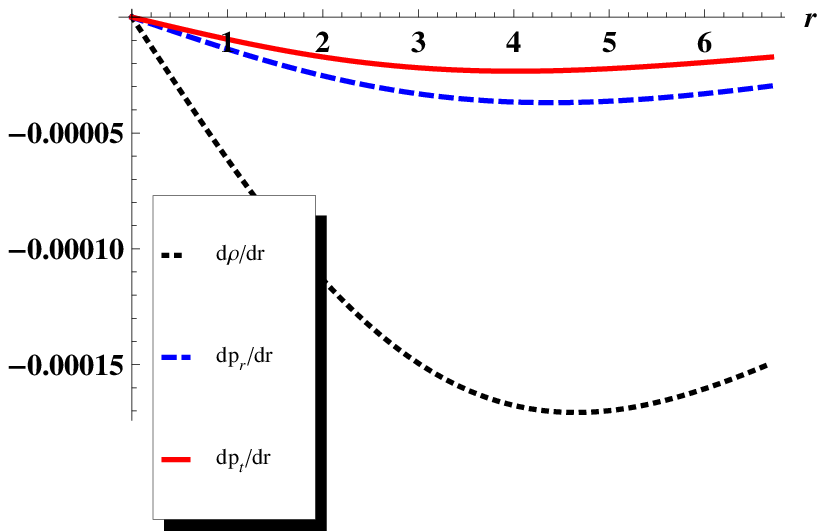}
       \end{minipage}
   \caption{(Left) The equation of state parameter $\frac{p_r}{\rho}$ (dashed line in blue color) and $\frac{p_t}{\rho}$ (solid line in red color) are plotted against $r$ for the compact star Her X-1 by taking the same values of the constant mentioned in Fig. 1. (Right) $\frac{d\rho}{dr}$ (dotted line in black), $\frac{dp_r}{dr}$ dashed line in blue and $\frac{dp_t}{dr}$ (solid line in red) plotted against $r$ for the compact star Her X-1 by taking the same values of the constant mentioned in Fig. 1} \label{f3}
\end{figure}

\section{Matching of interior and exterior spacetime}
Assuming the exterior spacetime to be the Schwarzschild exterior solution which is match smoothly with our interior solution and is given by
\begin{eqnarray}
ds^2 &=& \left(1-{2M\over r}\right) dt^2-\left(1-{2M\over r}\right)^{-1}dr^2-r^2(d\theta^2+\sin^2 \theta d\phi^2) \label{ext}
\end{eqnarray}

By matching the interior solution (\ref{metric}) and exterior solution (\ref{ext}) at the boundary $r=r_b$ we get
\begin{eqnarray}
e^{\nu_b} &=& 1-{2M \over r_b} = \left(A+\frac{B \left(b r_b^2+2\right)^{3/2}}{3 \sqrt{b}}\right)^2\label{bou1}\\
e^{-\lambda_b} &=& 1-{2M \over r_b} = (1+br_b^2)^{-2} \label{bou2}\\
p_r(r_b) &=& 0 \label{bou3}
\end{eqnarray}

Using the boundary condition (\ref{bou1}-\ref{bou3}),  we get
\begin{eqnarray}
\frac{A}{B} & = & \frac{2-4br_b^{2}-b^{2}r_b^{4}}{3\sqrt{b}\sqrt{2+b r_b^{2}}} \label{bou4}\\
A & = & (1+br_b^2)^{-1}-\frac{B \left(b r_b^2+2\right)^{3/2}}{3 \sqrt{b}} \label{bou5}\\
{2M \over r_b} &=& 1-(1+br_b^2)^{-2}\label{bou6}
\end{eqnarray}
Solving (\ref{bou4}-\ref{bou6})we get,
\begin{equation}
b=\frac{1}{r_b^{2}}\left[\frac{1}{\sqrt{1-\frac{2M}{r_b}}}-1\right]
\end{equation}
\begin{equation}
A=\frac{2-4br_b^{2}-b^{2}r_b^{4}}{6(1+br_b^{2})}
\end{equation}
\begin{equation}
B=\frac{\sqrt{b}}{2}\frac{\sqrt{2+br_b^{2}}}{1+br_b^{2}}
\end{equation}
Now the values of the constants $b,~A$ and $B$ for some well known compact stars are obtained in Table.1.\\

\begin{table}[h]
\tbl{The values of the constants $b$, $A$ and $B$ are obtained from the model for some well known compact star.}
{\begin{tabular}{@{}ccccccc@{}} \toprule
Compact star & M/$M_{\odot}$ & Radius(km) & b($km^{-2}$) & A & B($km^{-1}$) & $\frac{M}{r}$\\
 \colrule
RXJ 1856-37&0.9031 & 6 & 0.009475933306 & 0.06450929174 & 0.05552928200 & 0.2220120833\\
Her X-1 & 0.98&6.7 & 0.007268209067 & 0.07395029811 & 0.04902096388 & 0.2157462687\\
Vela X-12 & 1.77 & 9.99 & 0.004483071049 & 0.00117217295 & 0.03618423648 & 0.2613363363\\
Cen X-3 & 1.49 & 9.51 & 0.004020398660 & 0.05052417566 & 0.03574401806 & 0.2310988433\\
\botrule
\end{tabular} \label{ta1}}
\end{table}

\section{Conditions for well behaved solutions}

For well-behaved nature of the solutions for an an-isotropic fluid sphere the following conditions should be satisfied:

\begin{enumerate}
\item  The solution should be free from physical and geometric singularities, i.e. it should yield finite and positive values of the central pressure, central density and nonzero positive value of  $e^\nu|_{r=0}$ and  $e^\lambda|_{r=0}=1$.

\item  The causality condition should be obeyed i.e. velocity of sound should be less than that of light throughout the model. In addition to the above the velocity of sound should be decreasing towards the surface i.e.${d \over dr}~{dp_r \over d\rho}<0$  or   ${d^2 p_r\over d\rho^2}>0$   and ${d\over dr}{dp_t \over d\rho}<0$  or   ${d^2p_t \over d\rho^2}>0$  for $0\leq r\leq r_b$ i.e. the velocity of sound is increasing with the increase of density and it should be decreasing outwards.

\item  	The adiabatic index,  $\gamma= {\rho+p_r \over p_r}{dp_r \over d\rho}$  for realistic matter should be  $> 4/3$.

\item   The anisotropy factor $\Delta$ should be zero at the center and increasing outward.

\item   For a stable anisotropic compact star, $-1 \le v_t^2-v_r^2 \leq 0$ must be satisfied, \cite{herrera97} .

\end{enumerate}
We will check all the conditions one by one in the coming sections.

\section{Properties of the new solution}
The central pressure and density at the interior is given by
\begin{eqnarray}
8\pi p_r(r=0) & = & 8\pi p_t(r=0)= b \left(\frac{6 \sqrt{2} B}{3 A \sqrt{b}+2 \sqrt{2} B}-2\right)>0\label{pc}\\
8\pi\rho(r=0) & = & 6b>0;~~\forall~b>0
\end{eqnarray}
Plugging $G$ and $c$ we have obtained central density, surface density and central pressure of some well known stars which is given in Table 2.\\

To satisfy Zeldovich's condition at the interior, $p_r/\rho$ at center must be $\le 1$. Therefore
\begin{eqnarray}
\frac{ \sqrt{2} B}{3 A \sqrt{b}+2 \sqrt{2} B}-{1 \over 3} \le 1 \label{zel}
\end{eqnarray}

On using (\ref{pc}) and (\ref{zel}) we get a constraint on $A/B$ given as
\begin{equation}
-{5\sqrt{2}\over 12\sqrt{b}} \le {A \over B} < {3\sqrt{b} \over \sqrt{2}}
\end{equation}

Now the pressure and density gradients can be written as
\begin{eqnarray}
8\pi {d\rho \over dr} & = & -\frac{2 b^2 r \left(b^2 r^4+4 b r^2+15\right)}{\left(b r^2+1\right)^4}\\
8\pi {dp_r \over dr} & = & \frac{2 b^2 r [f_1(r)+f_r(r)]}{\left(b r^2+1\right)^3 \sqrt{b r^2+2} \left(3 A \sqrt{b}+B \left(b r^2+2\right)^{3/2}\right)^2}\\
8\pi {dp_t \over dr} & = & -\frac{3 b^2 r [f_3(r)+f_4(r)]}{\left(b r^2+1\right)^4 \left(b r^2+2\right)^{3/2} \left(3 A \sqrt{b}+B \left(b r^2+2\right)^{3/2}\right)^2}
\end{eqnarray}

where
\begin{eqnarray}
f_1(r) & = & 9 A^2 b \sqrt{b r^2+2} \left(b r^2+3\right)+3 A \sqrt{b} B \left(2 b^3 r^6+14 b^2 r^4+23 b r^2+3\right)\\
f_2(r) & = & B^2 \sqrt{b r^2+2} \left(b^4 r^8+9 b^3 r^6+12 b^2 r^4-22 b r^2-36\right)\\
f_3(r) & = & -36 A^2 b \left(b r^2+2\right)^{3/2}-3 A \sqrt{b} B \left(5 b^3 r^6+32 b^2 r^4+55 b r^2+24\right)\\
f_4(r) & = & 2 B^2 \sqrt{b r^2+2} \left(b^4 r^8+4 b^3 r^6+11 b^2 r^4+24 b r^2+20\right)
\end{eqnarray}

\subsection{Energy condition}
For physically acceptability our proposed model of compact star should satisfy the null energy condition (NEC), weak energy condition (WEC), strong energy condition (SEC) if the following inequalities hold at every points in the interior of a star.
\begin{eqnarray}
\text{NEC}: \rho(r)-p_r & \geq &  0\\
\text{WEC}: \rho(r)-p_r(r) & \geq &  0~~ \text{and} ~~\rho \geq  0\\
\text{SEC}: \rho(r)-p_r(r) & \geq &  0~~ \text{and} ~~\rho-p_r(r)-2p_t(r) \geq  0\\
\end{eqnarray}

\begin{figure}[!htb]\centering
   \begin{minipage}{0.55\textwidth}
     \includegraphics[width=\linewidth]{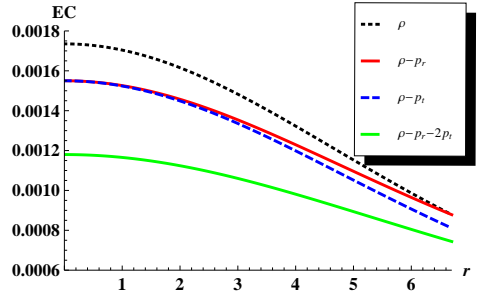}
       \end{minipage}
   \caption{The NEC, WEC and SEC are plotted against $r$ for the compact star Her X-1 by taking the same values of the constant mentioned in Fig. 1} \label{f4}
\end{figure}

Due to the complexity of the expression we will take the help of graphical representation. The LHS of the above inequalities are plotted in Fig. \ref{f4}, which implies that the above energy conditions proposed above are satisfied by our model of compact star.

\subsection{Mass-radius relationship}
The mass function of the solution can be determined using the equation given below
\begin{eqnarray}
e^{-\lambda} & = & 1-{2m \over r}\label{m1}
\end{eqnarray}
On using (\ref{m1}) in (\ref{g3a}) we get
\begin{eqnarray}
m(r) & = & 4\pi \int_0^r \rho r^2~ dr=\frac{1}{2}  \left(r-\frac{r}{ \left(b r^2+1\right)^2}\right)
\end{eqnarray}

\begin{figure}[!htb]\centering
   \begin{minipage}{0.49\textwidth}
     \includegraphics[width=\linewidth]{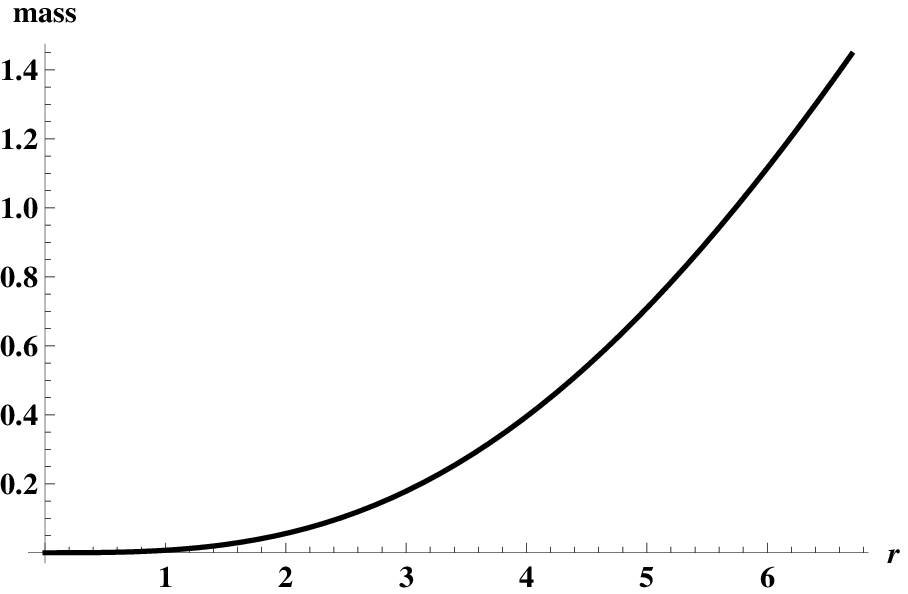}
        \end{minipage}
   \begin{minipage}{0.49\textwidth}
      \includegraphics[width=\linewidth]{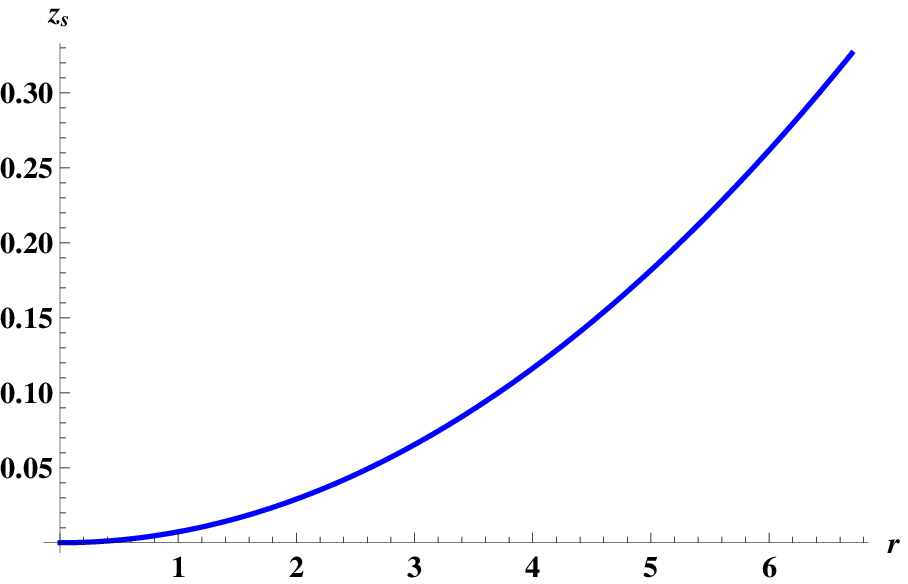}
       \end{minipage}
   \caption{(Left) The mass function is plotted against $r$ for the compact star Her X-1 by taking the same values of the constant mentioned in Fig. 1.  The surface red-shift is plotted against $r$ for the compact star Her X-1 by taking the same values of the constant mentioned in Fig. 1} \label{f5}
\end{figure}

The profile of mass function are shown in Fig. \ref{f5}. Figure indicates that mass function is a monotonic increasing function of $r$ and $m(r)>0$ everywhere within the boundary except at center.\\

For a model of compact star the ratio of mass to the radius can not be arbitrarily large, it should lie in the range $\frac{2M}{r}<\frac{8}{9}$, proposed by Buchdahl \cite{buchdahl59} . We have also calculated the value of $\frac{M}{r}$ for different compact stars from our model which is shown in Table 1. One can easily check that Buchdahl's condition is satisfied by our model of compact star.\\

The surface red-shift can be determine by
\begin{eqnarray}
z_s & = & e^{-\nu_b/2}-1=\left(A+\frac{B \left(b r^2+2\right)^{3/2}}{3 \sqrt{b}}\right)^{-1}-1
\end{eqnarray}
We have also drawn the profile of the surface red-shift which is shown in Fig. \ref{f5}. The Maximum value of the surface red-shift of some compact stars are obtained in Table 2. The table shows that the value of the surface redshift $z_s \leq 2$, lies in the expected range of Barraco \& Hamity \cite{hamity} .

\section{Stability Analysis of the model}
\subsection{Causality Condition}
A model of anisotropic compact star will be physically acceptable if the radial and transverse velocity of sound will be less than 1, known as causality conditions. The radial velocity $(v_{sr}^{2})$ and transverse velocity $(v_{st}^{2})$ of sound can be obtained as
\begin{equation}
v_{sr}^2={dp_r/dr \over d\rho/dr},~~~v_{st} ^2={dp_t/dr \over d\rho/dr}
\end{equation}

\begin{figure}[!htb]\centering
   \begin {minipage}{0.55\textwidth}
     \includegraphics[width=\linewidth]{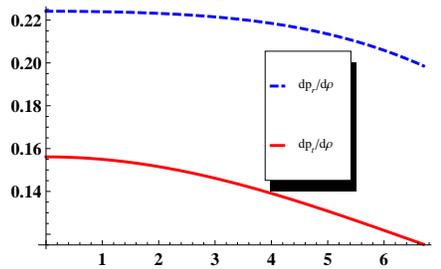}
      \end{minipage}
   \caption{Variation of square of radial and transverse velocity of sound are plotted against $r$ for the compact star Her X-1 by taking the same values of the constant mentioned in Fig. 1}\label{f6}
\end{figure}

\begin{figure}[!htb]\centering
   \begin{minipage}{0.49\textwidth}
     \includegraphics[width=\linewidth]{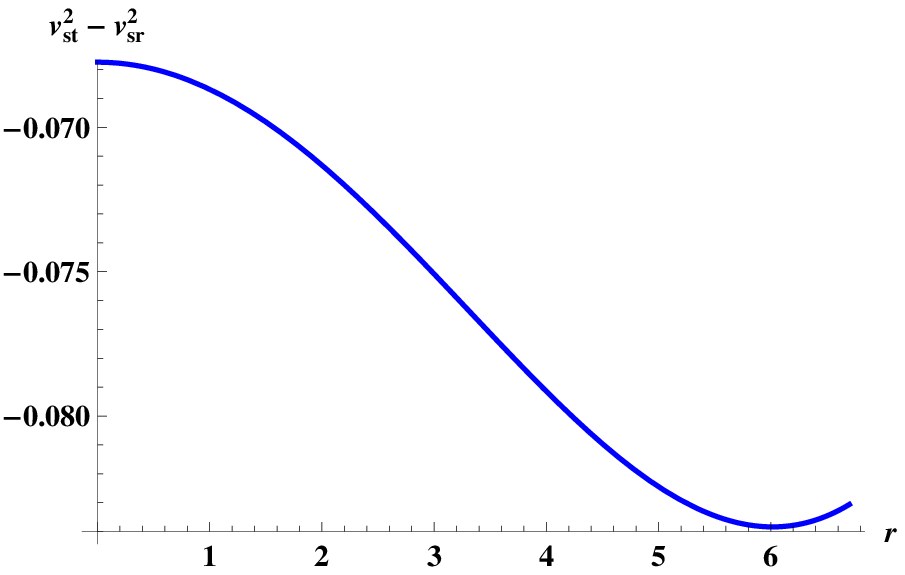}
      \end{minipage}
   \begin {minipage}{0.49\textwidth}
     \includegraphics[width=\linewidth]{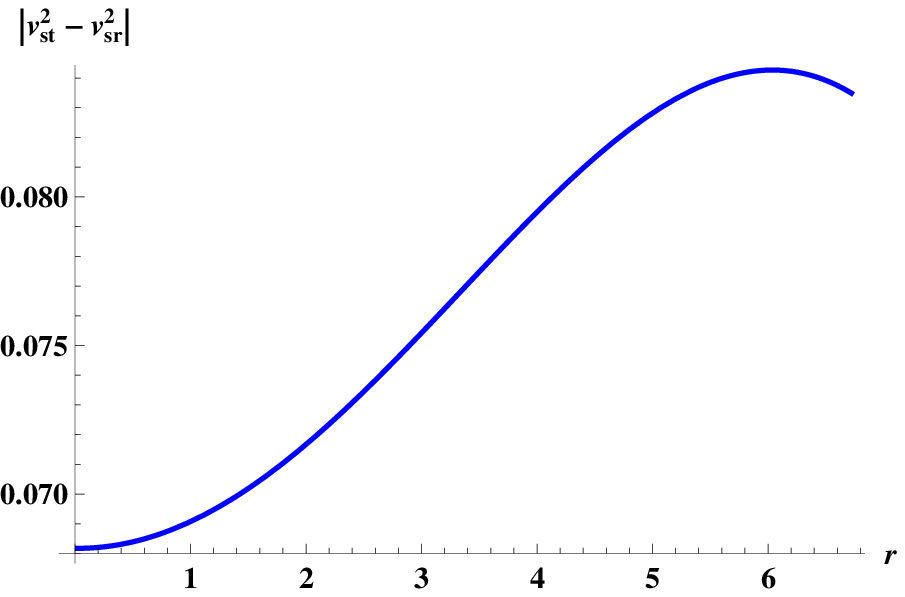}
       \end{minipage}
   \caption{(Left) Variation of $v_t^2-v_r^2$ with radius for Her X-1. (Right)  Variation of $|v_t^2-v_r^2|$ with radius for Her X-1.}\label{f7}
\end{figure}

The profile of radial and transverse velocity of sound have been plotted in Fig. \ref{f6}, the profile that our shows that causality condition is satisfied by our model. Using the concept of ``cracking" proposed by Herrera \cite{herrera92} to study the stability of anisotropic stars under the radial perturbations, Abreu {\em et al.} \cite{ab} proved that the region of an anisotropic fluid sphere where $-1\leq v_{st}^{2}-v_{sr}^{2} \leq 0$ is potentially stable but the region is potentially unstable where $0 \leq v_{st}^{2}-v_{sr}^{2}\leq 1$. From Fig. \ref{f7} it is clear that our model is potentially stable. Moreover for an anisotropic model of compact star $0<|v_{st}^{2}-v_{sr}^{2}|<1$ according to Andr\'{e}asson \cite{andreson} . Fig. \ref{f7} indicates that cracking method and Andr\'{e}asson's condition are verified.\\

\subsection{Stability under three forces acting on the system}
We want to examine the stability of our present model under three different forces $viz$ gravitational force, hydrostatics force and anisotropic force which can be described by the following equation
\begin{equation}
-\frac{M_G(r)(\rho+p_r)}{r}e^{\frac{\nu-\lambda}{2}}-\frac{dp_r}{dr}+\frac{2}{r}(p_t-p_r) = 0, \label{tov2}
\end{equation}
proposed by Tolman-Oppenheimer-Volkov and named as TOV equation.

The quantity $M_G(r) $ represents the gravitational mass within the radius $r$, which can derived from the Tolman-Whittaker formula and the Einstein's field equations and is defined by
\begin{equation}
M_G(r)=\frac{\nu'}{2}re^{\frac{\lambda-\nu}{2}}
\end{equation}
Plugging the value of $M_G(r)$ in equation (\ref{tov2}), we get
\begin{equation}
-\frac{\nu'}{2}(\rho+p_r)-\frac{dp_r}{dr}+\frac{2}{r}(p_t-p_r)=0.
\end{equation}
The above expression may also be written as
\begin{equation}
F_g+F_h+F_a=0,
\end{equation}

\begin{figure}[!htb]\centering
   \begin {minipage}{0.55\textwidth}
     \includegraphics[width=\linewidth]{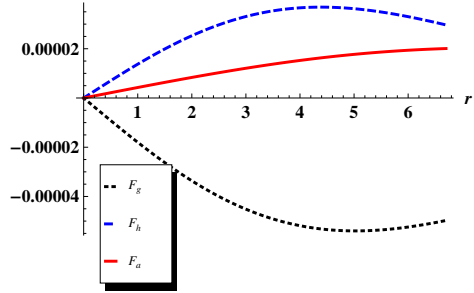}
       \end{minipage}
   \caption{Counter-balancing of three forces in TOV equation for static and equilibrium configuration for Her X-1.}\label{f8}
\end{figure}

Here
\begin{eqnarray}
F_g & = & -\frac{\nu'}{2}(\rho+p_r)\\
F_h &=& -{dp_r \over dr}\\
F_a &=& {2\Delta \over r}\\
\end{eqnarray}
The three different forces acting on the system are shown in Fig. \ref{f8}. The figure shows that gravitational force is negative and dominating in nature which is counterbalanced by the combine effect of hydrostatics and anisotropic forces to keep the system in equilibrium.

\subsection{Adiabetic Index}
For any model of compact star the adiabatic index $\Gamma$ must be always greater than $4/3$ for static equilibrium according to Heintzmann and Hillebrandt's \cite{hein} concept. The relativistic adiabatic index $\Gamma$ is given by
\begin{equation}
\Gamma = {\rho+p_r \over p_r}~{dp_r \over d\rho}
\end{equation}

\begin{figure}[!htb]\centering
   \begin {minipage}{0.5\textwidth}
     \includegraphics[width=\linewidth]{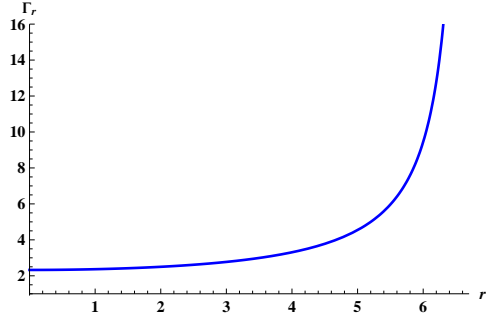}
       \end{minipage}
   \caption{Variation of adiabatic index $\Gamma$ with radius for Her X-1.}     \label{f9}
\end{figure}

The profile of $\Gamma$ are drawn in Fig. \ref{f9}. The figure shows that $\Gamma>\frac{4}{3}$ everywhere within the fluid sphere.

\begin{table}[h]
\tbl{The numerical values of central density, surface density, central pressure and surface red-shift are obtained for some well known compact star.}
{\begin{tabular}{@{}ccccccc@{}} \toprule
compact star & Central density $(\rho_c)$ & Surface Density ($\rho_b$) & Central Pressure & Surface Redshift\\
& (gm/cc.)&(gm/cc.) & ($dyne/cm^{2}$)\\
\colrule
RXJ 1856-37 & $3.054008814\times 10^{15}$ & $1.506563034\times10^{15}$ & $3.109131168\times 10^{35}$ & 0.341133599\\
Her X-1 & $2.342478977\times 10^{15}$ & $1.185726720\times10^{15}$ &$2.248396644\times 10^{35}$ & 0.326269905\\
Vela X-12 & $1.444853827\times10^{15}$ & $0.598972951\times10^{15}$ &$2.152357103\times 10^{35}$ & 0.447410939\\
Cen X-3 & $1.295738641\times 10^{15}$ & $0.615204142\times 10^{15}$ &$1.437436185\times 10^{35}$ & 0.363605257\\
\botrule
\end{tabular} \label{ta1}}
\end{table}

\section{Discussion}

Here we have successfully proposed a new model of anisotropic compact star in embedding class I spacetime  using a new type of  $g_{rr}$ metric potential. In this solution, all the physical quantities $\big(p_r,~p_t,~p_r/\rho,~p_t/\rho,~\rho,~v_r^2,~v_t^2 \big)$  are monotonically decreasing outward (Figs. \ref{f1}-\ref{f3}, \ref{f6}) and free from central singularities. Furthermore, the metric potentials, $\Delta,~\Gamma,~m(r)$ and $z_s$ are increasing function with the increase of radius (Figs. \ref{f1}, \ref{f2}, \ref{f5}, \ref{f9}). The decreasing nature of pressure and density is again reconfirmed by the negativity of their gradients Fig. \ref{f3}.

Our presented solution does satisfy the WEC, SEC and NEC (Fig. \ref{f4}) as well. The stability conditions are satisfied i.e. $-1\leq v_t^2-v_r^2 \leq 0$ and $0\leq |v_r^2-v_t^2|\leq 1$ (Fig. \ref{f7}). Therefore, our solution gives us the stable star configuration. Moreover, the TOV equation further supports the stability of the models by counter-balancing all the three forces each other to maintain the hydro-static equilibrium (Fig.  \ref{f8}).  The presented models of the above mentioned compact star candidates fit very well with the observed values of masses and radii. Hence our solution might have astrophysical relevance.

%\section*{Acknowledgments}

%----------------------------------------

\end{document}